\documentstyle[12pt]{article}
\begin{document}
\baselineskip 0.6cm
\begin{center}
{\bf
DENSITY-DEPENDENT SQUEEZING OF EXCITONS \\
IN HIGHLY EXCITED SEMICONDUCTORS}\\[0.5cm]
{\it by}\\
{\bf Nguyen Hong Quang}\footnote{On leave of absence from the
Institute of Theoretical Physics, P.O. Box 429, Hanoi 10000, Vietnam}\\
{\it International Center for Theoretical Physics}\\[1cm]
\end{center}

\begin{abstract}
The time evolution from coherent states to squeezed states of high
density excitons is studied theoretically based on the boson
formalism and within the Random Phase Approximation. Both the mutual
interaction between excitons and the anharmonic exciton-photon
interaction due to phase-space filling of excitons are included in
consideration. It is shown that the exciton squeezing depends
strongly on the exciton density in semiconductors and becomes smaller
with increasing the latter.
\end{abstract}

\vspace{2cm}
Recently the quantum optical properties of the quasiparticles in
condensed matters have been investigated intensively [1-10]. In
particular, the nonclassical properties of exciton-polariton were
considered in [1-3], phonon-polariton  in [4], and squeezed states of
excitons were considered in [5-7]. Of particular interest are also
the optical nonlinearities emerged from high external excitation
which affect almost all the optical properties of semiconductors.
At high excitation a great number of excitons can be produced and
there arise some kinds of anharmonicities such as nonlinear
exciton-exciton interaction and the anharmonic exciton-photon
interaction due to phase-space filling of excitons. An attempt to
consider the effect of these anharmonicities on squeezed states of
excitons was first made in [5-7]. However, in [5,6] to solve the
problem the authors had to resort to rough approximation neglecting
many terms in Hamiltonian which actually should be retained. It is
well known that for the high density systems where the fluctuations
of the occupation number can be neglected, the Random Phase
Approximation (RPA) [11] is the most suitable, and one can safely
apply it to study nonlinear optical properties of the system. In this
paper, as in [7], based on the boson formalism and within the
framework of RPA, we study the effect of both above-mentioned two
kinds of anharmonicities on the squeezing properties of high density
excitons. It will be shown that the squeezing degree of excitons
depends strongly on the exciton density in semiconductors, and moreover,
it decreases with increasing the latter.

We shall start with the following effective boson Hamiltonian
describing the exciton-photon system at high excitation [9]
\begin{equation}
H_{eff} = E^{ex}a^+a +\omega^\gamma c^+c + f a^+a^+ aa + [g_o c
a^+(1- \lambda a^+ a) + {\rm h.c.}] ,
\end{equation}
where $a^+ (a), c^+ (c)$ are the creation (annihilation) operators for
exciton and photon;\linebreak
$E^{ex}$, and $ \omega^\gamma $ are the energies of the exciton and photon,
respectively, $g_o$ is photon-exciton coupling constant.
The third term on the right-hand side of (1) describes the repulsive
exciton-exciton interaction, and $f$ is given by
\begin{equation}
f = - \frac{2}{V^3}\sum_{k,k'} V(k-k') \left[
|\varphi(k)|^2|\varphi(k')|^2 - |\varphi(k)|^2|\varphi(k)\varphi(k')|
\right] .
\end{equation}
The fourth term on the right-hand side of (1) represents the
anharmonic  exciton-photon interaction due to phase-space filling of
excitons, and $\lambda $ is given by
\begin{equation}
\lambda = \frac{1}{V} \sum_{k} |\varphi(k)|^2\varphi(k)/\sum_k
\varphi (k) .
\end{equation}
Here $V$ is the quatization volume, $\varphi(k)$ - is the Fourier
transform of the exciton envelope function for electron-hole relative
motion, $V(k-k')$ - is Coulomb potential.

We shall not detail here the derivation of Hamiltonian (1) reminding
only that it is obtained by operating Usui transformation [12] on the
Hamiltonian of the interacting photon-electron-hole system. This
procedure was originally developed by the authors of [13,14] for
boson formalism of high density exciton theory.

Let us consider the following Green's functions
\begin{eqnarray}
G_1(k,t) & = & \theta (t) < [a(t), a^+ (o) ] > \equiv << a(t) , a^+
(o) >>  ,\\[0.5cm]
G_2(k,t) & \equiv & << c(t) , a^+(o) >> .
\end{eqnarray}

The equations of motion for the Green's functions (4,5) can be
derived in standard way. Then using the RPA-type factorization which
is reasonable for high densities, we obtain the closed system of
equations, the solutions of which give us the spectrum of the
exciton-polariton, as follows

\begin{equation}
\epsilon_{1,2} = \frac{1}{2} (E^* + \omega^\gamma ) \pm \frac{1}{2}
\sqrt{{(E^* - \omega^\gamma)}^2 + 4 {g^*}^2} ,
\end{equation}
where
\begin{eqnarray}
E^* &=& E^{ex} + 2 n_{ex} V f , \\[0.5cm]
{g^*}^2 &=& g_o^2  ( 1 - n_{ex}V \lambda ) ,
\end{eqnarray}
and $n_{ex}$ is the density of excitons created by laser field in
semiconductors.

The polariton whose spectrum is (6) can be now expressed in terms
of photon  $c$ and ``real'' exciton $\tilde {a}$ operators, as follows
\begin{equation}
B_i = u_i c + v_i \tilde a ,
\end{equation}
where
\begin{eqnarray}
u_i &=& [1+ {g^*}^2 {(\epsilon_i - E^*)}^{-2}] ^{-1/2} , \\[0.5cm]
v_i &=& g^* |\epsilon_i - E^*| ^{-1} u_i .
\end{eqnarray}

To analyze the time evolution of the squeezing properties of the
exciton field we introduce two Hermitian quadrature exciton operators
\begin{eqnarray}
X(t) &=& \frac{1}{2} [ \tilde{a}^+(t) + \tilde{a}(t) ] , \\[0.5cm]
P(t) &=& \frac{i}{2} [ \tilde{a}^+(t) - \tilde{a}(t) ] .
\end{eqnarray}

{}From (9) and the exact time-dependent expression for operator
$B_i(t)$ one can obtain the following equation
\begin{equation}
\tilde{a} (t) = \sum_{i= 1,2 } \left[ u_i v_i c(o) + v_i^2 \tilde{a}(o) \right]
e^{\textstyle -i \epsilon_i t} .
\end{equation}
Since the squeezed states of exciton are defined (analogous to light)
as the states with a variance in one quadrature exciton operator
smaller than that associated with the coherent state, the exciton
becomes squeezed if
\begin{eqnarray}
<:(\Delta X(t))^2:> &=& <(\Delta X(t))^2> - \frac{1}{4} < 0 , \\[0.5cm]
{\rm or} \qquad <:(\Delta P(t))^2:> &=& <(\Delta P(t))^2> -
\frac{1}{4} < 0 ,
\end{eqnarray}
where $<:(\Delta X(t))^2:>$ and $ <:(\Delta P(t))^2:>$
are so-called the normally ordered variances and are widely accepted
as a measure of squeezing [15,16].

Let us, for simplicity, suppose that at initial time $ t= o$ excitons
are in coherent state and light is prepared in a squeezed state
characterized by the squeezed factor $ r$ and a phase $\varphi$ [1].
{}From (9), (12)-(14) and after some algebraic manipulations we obtain
the following exact analytical expressions for the exciton normally
ordered variances
\begin{eqnarray}
<:(\Delta X(t))^2:> &=& \frac{1}{2} |\beta '(t)|^2 \sinh^2 r -
\frac{1}{2} {\rm Re}(\beta '^2(t) e^{i\varphi})\cosh r \sinh r ,\\[0.5cm]
<:(\Delta P(t))^2:> &=& \frac{1}{2} |\beta '(t)|^2 \sinh^2 r +
\frac{1}{2} {\rm Re}(\beta '^2(t) e^{i\varphi})\cosh r \sinh r ,
\end{eqnarray}
where
\begin{eqnarray}
|\beta'(t)|^2 &=& 4 u_1^2 v_1^2 \sin^2 \frac{(\epsilon_1 -
\epsilon_2)t}{2} , \\[0.5cm]
{\rm Re}(\beta '^2(t) e^{i\varphi }) &=& -4 u_1^2 v_1^2 \sin^2
\frac{(\epsilon_1 - \epsilon_2)t}{2} \cos(\epsilon_1 + \epsilon_2 -
\varphi)t .
\end{eqnarray}
where $\epsilon_1, \epsilon_2$ and $u_1, v_1$  are defined above by (6)-(8) and
(10),(11), respectively.

The formulas (17)-(20) are our main results from which
we would like to emphasize the followings.
Since $u_i, v_i$ are expressed in  $ \epsilon_i , E^*$ and $g^*$
(see (10),(11)) which in turn depend explicitly on light frequency
$\omega$ , coupling constants $f, \lambda $ , and the exciton density
$n_{ex}$ (see (6)-(8)), one can expect various kinds of dependence of
the exciton squeezing on these parameters.

Note that the exciton density $n_{ex}$ always comes along with the
parameters $f$ and $\gamma$ characterizing the nonlinear
exciton-exciton interaction and anharmonicity of exciton-photon
coupling due to phase space filling of exciton. The coupling
constants $f$ and $\gamma$ were calculated for the lowest 1S-exciton
in bulk semiconductors in [9,14]
\begin{equation}
f =\frac{13\pi}{3V} E_Ba^3_B \ \ ,\qquad \gamma = \frac{7\pi}{2V} a^3_B .
\end{equation}
where $a_B$ is the effective Bohr radius of the bulk exciton and
$E_B$ is its binding energy.

For graphical illustration we take the case of CdS with the
parameters $E^{ex} = 2553 $mev , $ g_o = 50 $mev , $a_B =
2.8\times 10^{-7} $cm , $E_B = 27.8 $mev and $\varphi = 0 , r = 0.8 $,
scaled time is equal to $E^{ex} t$. To see the frequency dependence of
the exciton squeezing, in Fig. 1a, b, c we plot the time evolution of
the exciton normally ordered variance $<:(\Delta x(t))^2: >$ for the
fixed exciton density $n_{ex} = 1.10^{17}$ cm$^{-3}$, and three different
values of photon energy $\Delta \omega = E^{ex} - \omega = 350 $meV, 100
meV and 0 meV, respectively. One can see that with photon frequency
approaching closer to the exciton resonance ($\Delta \omega \rightarrow
0$) the squeezing degree obtained by excitons from photons effectively
increases, and gains its maximum at exact exciton resonance (Fig. 1c).

One of the new main results in this paper is explicit dependence of
exciton squeezing on the exciton density. This is a consequence of the
effects of nonlinear interaction between excitons and the anharmonic
exciton-photon interaction due to phase space filling of excitons. Higher
exciton densities more screen the interaction between excitons and
photons, and reduce it by the factor ($1 - n_{ex}V\lambda$). This is just
the saturation effect due to phase space filling of excitons which
usually appears in highly excited systems. The above-analyzed density
dependence of exciton squeezing is well seen from Fig. 2a, b, c plotted
for the fixed photon frequency near exciton resonance $\Delta\omega = 3$
meV, and some different exciton densities $n_{ex}$.\\[1.5cm]
{\bf Acknowledgements}

We would like to take this opportunity to express our gratitude to Prof.
Abdus Salam, the International Atomic Energy Agency and UNESCO for
hospitality at the International Centre for Theoretical Physics, Trieste.
We also would like to thank Prof. P.N. Butcher and Prof. Yu Lu for their
encouragements and discussions.\\[1.5cm]
{\bf REFERENCES}
\begin{enumerate}
\baselineskip=0.5cm
\item M.Artoni and J.L.Birman, Quantum Opt. {\bf 1} (1989) 91;
Phys.Rev.B{\bf 44} (1991) 3736 ; Opt.Commun. {\bf 89} (1992) 324;
{\bf 104} (1994) 319
\item V.R.Misko, S.A.Moskalenko, A.H.Rotaru and Yu.M.Shvera,
Phys.Stat.Sol(b) {\bf 159} (1990) 477
\item P.Schwendimann and A.Quattropani, Europhys.Lett. {\bf 17}
(1992) 355
\item A.V.Chizhov, B.B.Govorkov, A.S.Shumovsky, Mod.Phys.Lett.B {\bf
7} (1993) 1233
\item N.B.An, J.Phys.Condens.Matter {\bf 5} (1993) 8347
\item N.B.An and T.T.Hoa Phys.Lett.A {\bf 180} (1993) 145
\item N.H.Quang, Proc. of IV Nat. Conf. on Phys. 5--8/10/1993, pp. 155--160
\item S.Schmitt-Rink, D.S.Chemla and D.A.B.Miller, Adv.Phys.{\bf 38} (1989) 89
\item T.Hiroshima, Phys.Rev.B{\bf 40} (1989) 3862 ;
J.Phys.Condens.Matter {\bf 4} (1992) 3849
\item N.M.Khue, N.Q.Huong, N.H.Quang, J.Phys.Condens.Matter {\bf 6}
(1994) 3221
\item N.N.Bogolubov and S.V.Tjiablikov, Sov.Phys.Doklady {\bf 4} (1959) 589
\item T.Usui, Prog.Theor.Phys.{\bf 23} (1960) 787
\item E.Hanamura, J.Phys.Soc.Jpn.{\bf 29} (1970) 50 ; {\bf 37} (1974)
1545, 1553
\item  E.Hanamura and H.Haug, Phys.Rep.{\bf 33C} (1977) 209
\item S.M.Barnett, Opt.Commun. {\bf 61} (1987) 432
\item V.Bu\u{z}ek and I.Jex, Intern. J.Mod.Phys. B {\bf 4} (1990) 659
\end{enumerate}
\newpage

{\bf Figure Captions}\\[2cm]
{\bf Fig.1} Time evolution of the exciton normally ordered variance
$<:(\Delta x(t))^2: >$ for $n_{ex} = 1.10^{17}$ cm$^{-3}$; $\Delta \omega
= 350 $meV (a), 100 meV (b) and 0 meV (c). The closer to exciton resonance
the greater exciton squeezing achieved.\\[2cm]
{\bf Fig.2} Time evolution of the exciton normally ordered variance
$<:(\Delta x(t))^2: >$ for  $\Delta \omega = 3 $meV;
$n_{ex} = 1.10^{17}$ cm$^{-3}$ (a), 1.10$^{18}$ cm$^{-3}$ (b) and
3.10$^{18}$ cm$^{-3}$ (c). With increasing $n_{ex}$ the exciton squeezing
degree decreases.
\end{document}